\documentclass[11pt]{article}
\begin{document}
\begin{center}
\begin{Large} {\bf Circular Motion of a Small  Oscillator in a Zero-Point Field Without External Forces: Is It Possible?}
\end{Large}
\vspace{10mm} \\
\begin{Large}
{Yefim S. Levin }\\
Department of  Mathematics, Salem State University,
Salem MA, 01970   \\
\end{Large}
\end{center} 
\begin{abstract}
\indent A small dipole oscillator moving along a circular trajectory in zero-point
electromagnetic field (ZPF) and with a polarization normal to the rotation plane,  
 is considered.  Temporal periodicity conditions are imposed on ZPF, associated with the way the rotating oscillator ``observes" ZPF. They are similar to spatial boundary conditions in Casimir phenomenon  and therefore result in  ZPF spectrum change from continuous one to a discrete one and, as a consequence, an effective temperature of the modified ZPF \cite{ylevin2010}.  The average centripetal average force on the oscillator, originating from this modified ZPF scattered by the oscillator in the near zone, is calculated in terms of the bilinear correlation functions of electromagnetic field. After renormalization of the correlation function, which physically means extraction of a pure effect of periodicity, the force has a finite value. All calculations are carried out using the methodology  of stochastic electrodynamics. The radial component of the force is directed to the center of rotation.
In non relativistic case and for oscillator frequency smaller than rotation frequency, the force turns out to be proportional to the rotation radius. Such  result could mean a possibility that micro motion of the oscillator in ZPF  sustains its average circular motion  without any other external forces.Though the estimation done for the point-like electron shows that the effect is not observable because the radius of such circular electron motion would have been much smaller than the classical electron radius,  $R \ll r_{cl}$, when  our  semi classical approach to the electromagnetic problem does not work. It is well expected result and considered as the first step in the application of the idea in the quark world governed by non abelian colored fields, subject of the next paper. The problem under discussion can  be associated conceptually with the self-consistent approach to the particle-field problems \cite{Johnoson_Hu2001}, when the field configuration and particle motion are adjusted to each other. 
\end{abstract}
\section{Introduction.}     
\indent This work was inspired by two reasons. The first one can be addressed to the way how thermal effects associated with uniform acceleration of a detector in a vacuum are considered, both in quantum and classical theories \cite{birell1982}, \cite{boyer1984}. These approaches lead to the similar conclusions but there are some discrepancies between them to be expected which would be useful to explore. \\
\indent The widely accepted in the literature description  of a warmed radiation observed by a quantum detector uniformly  accelerating through a  vacuum is explicitly based on the assumption that the detector at a quantum state  moves along a classical trajectory.  It is obviously in contradiction to Heisenberg's Uncertainty Principle. On the other hand, in the classical approach   there is no such problem. The detector is envisioned as an uniformly accelerated classical oscillator and the classical trajectory is assigned to the oscillator center or, in other words, to an average location of the oscillator. So we could expect the same situation to be in the quantum case - distinguishing  `` instantaneous " and average motions of the oscillator. In quantum language it  means that an average location of the quantum oscillator should be associated with its  expectation value and therefore a ``classical trajectory " of the quantum detector can not be selected arbitrary, without connection with its inner quantum states. In other words, in any consistent quantum approach a ``classical trajectory" of the oscillator and its quantum state should be self-consistent, which is probably not easy to achieve
\cite{Johnoson_Hu2001}. Returning to the classical case \cite{boyer1984} of uniformly accelerating oscillator we see that there is no such strong connection between the motion of the oscillator and its center.
Nevertheless, as the first step to achieve self-consistency  mentioned above for the quantum case, it would be interesting to find a classical configuration when such self-consistency  exists. It means that a motion of the classical object ( electric charge or  oscillator) through the vacuum ( also described in classical terms ) near some trajectory changes the vacuum the way that mechanical reactions arising from 
the interaction with this modified vacuum  keep an average location of the  object on  this trajectory. In other words, the system consisting of  particle and  field vacuum  should be self-consistent. \\
\indent The second reason is associated with our curiosity:  can such classical self-consistency idea be extended to 
the quark-gluon world?  With the classical Wong equations \cite{Wong1970} in hand one can describe quark or, more exactly, color particle trajectory. However the concept of classical vacuum  of   color field in the Wong theory is not present. \\
\indent The Wong classical approach was developed at the time when quantization of the Yang-Mills equations 
was not completed. Now, with completed theory of quantization, from the quantum field theory of gauge fields, one can see that at some physical situations the classical picture represented by the Wong equations can be supplemented by the concept of ( let us call it ) Stochastic Gluon Vacuum ( SGV ). Indeed, at small distances , when running coupling constant is very small ( asymptotic freedom ) and perturbation theory works well, one can expect the situation in Quantum Chromodynamics ( QCD ) very similar to the Quantum Electrodynamics ( QED ) and its classical analog -
Stochastic Electrodynamics ( SED ). The SQV, in full analogy with SED, can be described as a set of all possible
plane gluon waves with random phases. Then self-consistent trajectories in both SED and Wong/SGV theory can be calculated in a similar way. \\
\indent In this work we investigate a self-consistent SED model even though we expect some unphysical ( not observable ) results. We consider this work as  the first step to a more `` realistic " model in Wong/SGV theory, the subject of the next paper.\\         
\indent The system we discuss in this work consists of two parts, a dipole oscillator and stochastic zero-point electromagnetic field (stochastic vacuum). Interaction between these parts changes both 
the oscillator motion and the vacuum. We assume that these changes can result in appearance of a stable on the 
average configuration. The most likely candidate for such stable average configuration is the oscillator moving
along an average circular orbit in the ZPF with a (modified) discrete spectrum. This system is assumed to be 
self-consistent. The oscillator, due to its rotation, observes the vacuum discrete spectrum. The mechanical reaction 
from the modified vacuum is an average force keeping the oscillator on the circular orbit. The  calculation of this average vacuum force is a central point of our work. An analogy of this force with the Casimir force can be 
be useful to understand its origin.\\
\indent  The Casimir force is considered as a result of finite and observable changes in the infinite zero-point energy of the vacuum which are connected with spatial boundary conditions.  In this work we consider vacuum forces, also  associated with vacuum fluctuations change, which however 
are due to ``temporal boundaries " or periodicity.  For illustration of this change, let us consider gedanken measurements of a two-point correlation function
of the  vacuum field made by a point-like observer uniformly  rotating about some point in the vacuum. The correlation
function depends on
the  difference of proper times of these two points, $\tau_2-\tau_1$, because of stationarity of the vacuum field. Moreover,  we have to expect the results of such measurements, made by the observer participating in a periodic motion, to be the same for any loop, and the correlation function should satisfy
periodic condition that is  $CF(\tau_2-\tau_1) = CF((\tau_2- \tau_1) + n \times T )$ ,where  $n=\pm 1,2,3,...$ and T is a period of rotation.  As a result of this periodicity condition
the observed spectrum of the vacuum field correlation function should contain  harmonics with frequencies multiple to the angular velocity of rotation, the effect well known from Fourier series theory, and  differ from the original vacuum continuous spectrum.  Below we will derive an expression for such a force exerting on a small z-oscillator uniformly rotating in xy-plane in the vacuum represented as a classic  zero-point  electromagnetic field.
Due to the vacuum fluctuations change, the vacuum force experienced by the rotating object has ( after renormalization) a finite value. \\
\indent According to our calculations the force has the component which is directed to a rotation center and could contribute to a net centripetal force keeping the moving object on a circle. Then there is a great temptation to see if  this vacuum force alone,  without any other external forces, can support this rotation? On this way we obtain relationship between the mass and angular velocity of the oscillator stationary rotating in the vacuum without any externally applied forces. It contains the Plank's constant and therefore has a quantum character, even though the computation is done in the frame of classical stochastic electrodynamics. It is estimated  for the case of non relativistic rotation.\\
\indent Speaking about self-consistency, we talked above,  one has to mention the approach \cite{Johnoson_Hu2001}, \cite{Lin_Hu2006} (see also references) which did not effect our work but conceptually   
is very closely related with ours. The approach is developed as a stochastic theory of relativistic particles moving in a quantum field. Even though the quantum field theory is the starting point of this approach, a particle in the field environment can participate in different regimes of the motion: quantum, stochastic, and semi classical.
The main philosophy of the approach is that a classical trajectory can not be prescribed, it has to be determined. 
\\
\indent We consider here only a steady state situation and choose a classical trajectory of a particle  which is self-consistent with the zero point field in which it moves and which it modifies.     
\section{The Average Force Calculation Method.}
\indent The average non zero net force exerted on an oscillator should be an expression bilinear in terms of an
applied stochastic electromagnetic field, $(\vec{E}_a, \vec{B}_a)$, because a linear expression of such field vanishes after averaging.
So it is different from the Lorentz force. To find  such force  we follow the approach developed in \cite{ford1998}.\\
\indent The force can be calculated by integrating the Maxwell stress tensor over the  spherical surface just outside the particle,
\begin{eqnarray}
F^{(i)}=\oint da_{(j)}T^{(ij)},
\end{eqnarray}   
where 
\begin{eqnarray}
T^{(ij)} =\frac{1}{4 \pi}  (  E^{(i)} E^{(j)}  +B^{(i)} B^{(j)}  -(1/2) \delta^{ij} ( \vec{E}^2 +\vec{B}^2   )   ),
\end{eqnarray}
\indent The net fields in this expression are  $\vec{E}= \vec{E}_a + \vec{E}_d $ and $\vec{B}= \vec{B}_a +
\vec{B}_d$, where $(\vec{E}_a, \vec{B}_a)$ is an applied electromagnetic field defined below and $(\vec{E}_d,\vec{B}_d)$ is the field scattered by the particle interacting with the applied field. The integration surface is taken very closely to the 
particle. So the scattered field is assumed to be taken in the near zone in dipole approximation: 
\begin{eqnarray}
\vec{E}_d \approx \frac{3 \hat{n}(\hat{n}\vec{p}-\vec{p})}{r^3}, &  \vec{B}_d \approx -\frac{\hat{n} 
	\times \vec{p}}{r^2}.
\end{eqnarray}  {\tiny }
Here r is the radial distance from the particle, $\vec{p}$ is its time-varying dipole moment induced by the applied field
and to be defined below , and $\hat{n}$ is the outward directed unit normal vector on the sphere surrounding the particle.\\ 
\indent Only the cross terms in $T^{(ij)}$ are of interest. So
 \begin{eqnarray}
 \vec{F}= \frac{1}{4 \pi} \oint da [(\hat{n}\vec{E}_a)\vec{E}_d + (\hat{n}\vec{E}_d)\vec{E}_a + (\hat{n}\vec{B}_a)\vec{B}_d - \hat{n}(\vec{E}_a\vec{E}_d +  \vec{B}_a\vec{B}_d   ).      ]
 \end{eqnarray}
\indent  Because the particle and the radius of the integration surface are assumed to be small, $\vec{E}_a$ and $\vec{B}_a$ can be expanded in a Taylor series around $\vec{x}= \vec{x}_0$, the location of the particle. The leading non zero contributions to the force come from the zero-order in $\vec{B}_a$ and the first order term in $\vec{E}_a:$
\begin{eqnarray}
\vec{B}_a(\vec{x},t) \approx \vec{B}_0, \;\;\; E^{(i)}_a(\vec{x},t) \approx E^{(i)}_0 +r \hat{n}\nabla E^{(i)}_0.
\end{eqnarray} 
Then after the angular integration the force takes the form
\begin{eqnarray}
\label{eq:FordForce}
\tilde{F}^{(i)}=  \frac{2}{3}(\dot{\vec{p}} \times \vec{B}_{0})^{(i)}+
\frac{1}{3}p_{(j)}\partial^{(i)} E^{(j)}_{0} + \frac{2}{3}p^{(j)} \partial_{(j)}E^{(i)}_{0} \equiv F^{(i)} + F_1^{(i)}+ F_2^{(i)}. 
\end{eqnarray}
\indent  We are going to use the method to calculate the average force exerted on a small oscillator
$ \textit{moving} $ on a circle with radius R and angular velocity $\Omega_0$, through random classical radiation according to 
\begin{eqnarray}
\label{eq:OscillatorRotation}
\vec{R}=( R \cos \Omega_{0}t, R \sin \Omega_{0}t, 0).
\end{eqnarray} 
So the method, developed in \cite{ford1998} for a static situation, in our case should be  modified. \\
\indent In our case the force (\ref{eq:FordForce}), obtained for a body $\textit{at rest}$,  is computed in an instantaneous inertial
reference frame $I_{\tau}$, with the oscillator equilibrium point also $\textit{at rest}$ in it, and $I_{\tau}$
is defined by 4 vectors $\mu^i_{(k)}(\tau)$ of a Frenet-Serret orthogonal tetrad \cite{synge}(55):
 \begin{eqnarray}
 \label{eq:mu1234}
 \mu^i_{(1)}(\tau)=(\cos \alpha, \sin \alpha, 0,0), \nonumber \\
 \mu^i_{(2)}(\tau)=(-\gamma \sin \alpha, \gamma \cos \alpha,0, \beta \gamma),
  \nonumber \\
  \mu^i_{(3)}(\tau)=(0,0,1,0), \nonumber \\
   \mu^i_{(4)}(\tau)= \frac{U^i(\tau)}{c}=(-\beta \gamma \sin \alpha,\beta \gamma \cos \alpha,
  0, \gamma),
   \end{eqnarray}
   where $U^i(\tau)$ is a 4 vector velocity of the oscillator equilibrium point in the laboratory system  at its proper 
time $\tau$ ,   $\beta=
   v/c=\Omega_0 R /c, \gamma=(1-\beta^2)^{-1/2}$, and $\alpha= \Omega_0
   \gamma \tau$. Besides
   \begin{eqnarray}
    \mu^i_{(a)} \mu_{{(b)}i}=\eta_{(ab)}, \;\;
    \eta_{(ab)}=diag(1,1,1,-1), \;\; a,b =1,2,3,4, \;\;\; g_{ik}=diag(1,1,1,-1).
    \end{eqnarray} 
\indent  All quantities associated with instantaneous frames $I_{\tau}$ in (\ref{eq:FordForce}),dipole momentum $p^{(j)}$, zero-point electric $E_0^{(j)}$
and magnetic  $B_0^{(j)}$  fields ( they are described below ), are supplied with indexes in parentheses.
Particularly, 4-velocity of the equilibrium point in $I_{\tau}$ is 
\begin{eqnarray}
 U_{(a)}=\mu^i_{(a)}(\tau)U_i(\tau)= \mu^i_{(a)}(\tau) U^k(\tau) g_{ik}= (0,0,0,-c).
 \end{eqnarray}
So, indeed, in the reference system S, consisting of all instantaneous inertial reference frames $I_{\tau}$, the oscillator equilibrium point is at rest at any proper time  $\tau$ and
its spatial coordinates are $x^{(i)}_0=(0,0,0)$. In this reference system, the z-oscillator coordinates are $x^{(i)}= (0, 0, z(\tau))$, and dipole moment is $ p^{(i)}(\tau)= (0,0, ez(\tau))$.\\
\indent The expression for the force $F^{(i)}$ was found for an arbitrary applied 
electromagnetic field but in an electric-dipole approximation. It can be a 
problem to match this restriction with a random classical radiation field, which spreads over all frequencies from 0 to $\infty$, because electric-dipole approximation can be broken for high frequencies due to 
 very small wave lengths. So the final result should be checked regarding validity of the electric-dipole approximation. \\
\indent Three more comments regarding our calculations of the force should be added here.  First, all our calculations are done in the frame of the stochastic electrodynamics (SED). \footnote {There is an opinion that SED can not be alternative to QED \cite{milonni1994}. But nevertheless it is useful tool to get understanding or intuition  of some problems. Besides there can be reasons to consider SED not as an alternative
to QED but as a necessary and essential part of a full quantum theory. Indeed, Philip R. Johnson and B. L. Hu discussing in \cite{Johnoson_Hu2001}, p.4 
`` The view of the emergence of semi classical solutions as decoherent histories" formulated in \cite{Gell-MannHartle1993}  suggest to consider classical solutions of particle-field systems ``in the context of how classical solutions arise from the quantum realm".} 
 Second, it is known that in the frame of SED the random classic radiation fields applied in our model to the rotating 
oscillator  and therefore $\tilde{F}^{(i)}$ depend on random phases. The $\textit{observable}$ force 
exerted on the rotating oscillator can be obtained from $\tilde{F}^{(i)}$ only after averaging it over these random phases. It will be taken to be $ <\tilde{F}^{(i)}>$, where the symbol $ <...> $ means averaging.
 Third, we use rotating Frenet-Serret orthogonal tetrads (other than more often referred Fermi-Walker tetrads) because it is convenient that vector $\mu^i_{(1)}(\tau)$ of the tetrad is directed along the radius of rotation,  $\mu^i_{(2)}(\tau)$ is tangent to the rotation circle, and $\mu^i_{(3)}$ is perpendicular to the rotation plane at all times $\tau$.
\section{ Classical Electromagnetic Zero Point Radiation.}
\subsection{Applied Field at the Location of the Oscillator in the Laboratory System}
The applied field, which was referred above as $\vec{E}_a, \vec{B}_a$  with index a, through which the oscillator moves, is classical electromagnetic zero point radiation. It has a form  \cite{boyer1980} :
\begin{eqnarray}
\label{eq:Field_In_Lab}
\vec{E}(\vec{r},t)= \sum^2_{\lambda=1} \int d^3k
\hat{\epsilon}(\vec{k},\lambda)h_0(\omega)\cos[\vec{k}\vec{r}-\omega
t -\Theta(\vec{k},\lambda)], \nonumber \\
\vec{H}(\vec{r},t)=\sum^2_{\lambda=1} \int d^3k
[\hat{k},\hat{\epsilon}^(\vec{k},\lambda)]h_0(\omega)\cos[\vec{k}\vec{r}-\omega
t -\Theta(\vec{k},\lambda)]. \label{eq:ff1}
\end{eqnarray}
where the $\theta(\vec{k},\lambda)$ are random phases distributed
uniformly on the interval $(0,2\pi)$ and independently for each
wave vector $\vec{k}$ and polarization $\lambda$ of  a plane
wave,
\begin{eqnarray}
 \pi^2 h_0^2(\omega)=(1/2)\hbar \omega,
 \end{eqnarray}
\begin{eqnarray}
\label{eq: CosSinAverageContSpectrum}
\langle \cos\theta(\vec{k}_{1}
\lambda_1 )\cos\theta(\vec{k}_{2} \lambda_2 ) \rangle= \langle
\sin\theta(\vec{k}_{1} \lambda_1 )\sin\theta(\vec{k}_{2}
\lambda_2 )\rangle=  \frac{1}{2}\delta_{\lambda_1 \, \lambda_2} \,
\delta^3(\vec{k}_1-\vec{k}_2),\;\;\;
\langle \cos\theta(\vec{k}_{1}
\lambda_1 )\sin\theta(\vec{k}_{2} \lambda_2 ) \rangle=0, \nonumber \\
\end{eqnarray}
 \begin{eqnarray}
\label{eq:lambda1}
  \sum^2_{\lambda=1}\epsilon_i(\vec{k}\lambda
)\epsilon_j(\vec{k}\lambda )= \delta_{ij}-\hat{k}_i\hat{k}_j
\end{eqnarray}
\indent It is convenient to use vector $\vec{k}=(k_x,k_y,k_z)$  in spherical coordinates $\vec{k}=(k, \theta, \phi)$ to separate integration over k 
from integration over angles $\theta$ and $\phi$:
\begin{eqnarray}
\int dk_x \,dk_y \, dk_z \Rightarrow \int dk\;k^2\, \int d\theta d\phi \; \sin \theta
\end{eqnarray} 
So vectors $\epsilon(\vec{k},\lambda)$ and  $[\hat{k},\hat{\epsilon}(\vec{k},\lambda)]$ can also be expressed in terms of spherical k-coordinates $(k,\theta, \phi)$. \\
\indent To find the electric and magnetic fields at the location of the oscillator center (we referred to them as 
$( \vec{E}_0 ,\vec{B}_0 )$ ) one has to insert  (\ref{eq:OscillatorRotation}) into  (\ref{eq:ff1}). 
It is easy to see then that the electric and magnetic fields at the location of the oscillator center are functions of time only, t in a lab system and $\tau$ in $I_{\tau}$,  and do not depend on spatial coordinates.
Even the $\vec{R}$-function (\ref{eq:OscillatorRotation}) is a periodic in time, the obtained this way formal expressions for the fields are not periodic  because of the term $ \omega t$ in (\ref{eq:ff1}). \\
\indent  From observation point of view it does not look reasonable. The oscillator center motion is periodic and 
all oscillator observables  also should be periodic, with 
the period $T=2 \pi/ \Omega_0$. This time periodicity caused by the observation is similar to a spatial periodicity of the vacuum  
imposed by boundary conditions in Casimir effect. In other words, any observations, including the fields, made by the oscillator at time interval $0 < t < T$ should be identical with the measurements at interval $n T < t < (n+1) T$, where $n=0, \pm 1, \pm 2, ...$.       \\
\indent To achieve it , in the expressions for observable  zero-point fields $\vec{E}_0$ and $\vec{H}_0$   continuous variable $\omega$ in (\ref{eq:ff1}) should be  changed to a discrete one,
 $\omega_n=\Omega_0 n$, and  integration over $k=\omega / c$  to summation over $k_n=\omega_n/c$: 
\begin{eqnarray}
\int_{0}^{\infty} dk \; k^2 f(k) \Rightarrow  k_0 \;\sum_{n=0}^{\infty}\kappa_n^2\; f(k_n).
\end{eqnarray}     
Then  the electric and magnetic components of the zero-point radiation field at the oscillator center location, $ \textbf{modified by the observation process} $, become: 
\begin{eqnarray}
\vec{E}_0(t_i)= k_0\:\sum^{\infty}_{n=0} \sum^2_{\lambda=1}
\int do\,k^2_n\,\hat{\epsilon}(\hat{k},\lambda)\,h_0(\omega_n)\
\cos[\vec{k_n}\vec{r}(t_i)-\omega_n
t_i -\Theta(\vec{k_n},\lambda)], \nonumber \\
\vec{H}_0(t_i)=k_0\:\sum^{\infty}_{n=0}\sum^2_{\lambda=1}
\int do \,k^2_n \,
[\hat{k},\hat{\epsilon}(\hat{k},\lambda)]\,h_0(\omega_n)\,
\cos[\vec{k_n}\vec{r}(t_i)-\omega_n
t_i-\Theta(\vec{k_n},\lambda)], \nonumber \\
 \vec{k}_n=k_n \hat{k},
\;\; k_n=k_0\,n, \;\; k_0= \frac{\Omega_0}{c}, \;\; \omega_n=c
\,k_n, \;\;
 do= d\theta \, d\phi \, \sin\theta, \;  \nonumber \\
 \hat{k}=(\hat{k}_x,
\hat{k}_y, \hat{k}_z)=(\sin\theta \, \cos \phi, \, \sin\theta \,
\sin\phi, \,\cos\theta\,).
\label{eq:mff1}
\end{eqnarray}
We have introduced index $i=1,2$ to distinguish the fields at two different
times, $t_1$ and $t_2$, which is necessary for further calculations.
The unit vector $\hat{k}$ defines a direction of the wave vector.  \\
 The expressions for the average over random phases are transformed first
to spherical momentum space \cite{davydov1968}, then to the
case of discrete values of "k", and take the form:
\begin{eqnarray}
\label{eq: CosSinAverage1} \langle \cos\theta(\vec{k}_{n_1}
\lambda_1 )\cos\theta(\vec{k}_{n_2} \lambda_2 ) \rangle= \langle
\sin\theta(\vec{k}_{n_1} \lambda_1 )\sin\theta(\vec{k}_{n_2}
\lambda_2 )\rangle=  \frac{1}{2}\delta_{\lambda_1 \, \lambda_2} \,
\frac{2}{k_0(k_0 n_1)^2}\, \delta_{n_1 \,
n_2}\delta(\hat{k}_1-\hat{k}_2). 
\end{eqnarray}
The equation (\ref{eq:lambda1}) does not depend on n and has the same form for
 both continuous and discrete spectrum.
\subsection{ Applied Field at the Location of the Oscillator in an Instantaneous Frame.}
\label{subsection: FieldIn InstantaneousFrame}
The expressions (\ref{eq:mff1}) are obtained  in the laboratory system. In an instantaneous reference frame $I_{\tau}$, after Lorentz
transform, corresponding expressions for the fields  at the time $\tau_i=t_i/ \gamma, \;\; i=1,2$, and at the location $r^{(k)}=(0,0,0)$ 
 are as follows
\begin{eqnarray}
\label{eq:field_Instantaneous}
E_{(1)}(\tau_i)= E_1(\tau_i) \gamma \cos \alpha_i + E_2(\tau_i) \gamma \sin \alpha_i +H_3(\tau_i) \beta \gamma, \nonumber \\
E_{(2)}(\tau_i)= E_1 (\tau_i)(-\sin \alpha_i) + E_2(\tau_i) \cos \alpha_i,\nonumber \\
E_{(3)}(\tau_i)= E_3(\tau_i) \gamma - H_1(\tau_i) \beta \gamma \cos \alpha_i -H_2(\tau_i) \beta \gamma \sin \alpha_i, \nonumber \\
H_{(1)}(\tau_i)= H_1(\tau_i) \gamma \cos \alpha_i + H_2(\tau_i) \gamma \sin \alpha_i - E_3(\tau_i) \beta \gamma, \nonumber\\
H_{(2)}(\tau_i)=H_1(\tau_i) (- \sin \alpha_i ) + H_2(\tau_i) \cos \alpha_i, \nonumber \\
H_{(3)}(\tau_i)=H_3(\tau_i) \gamma + E_1(\tau_i) \beta \gamma \cos \alpha_i + E_2 (\tau_i)\beta \gamma \sin \alpha_i,
\end{eqnarray}
where 
\begin{eqnarray}
\label{eq: alpha}
\alpha_i = \Omega_0 \gamma \tau_i, \;\;\; i=1,2.
\end{eqnarray}
Quantities with an index in parenthesis are to be taken for the field components in instantaneous inertial reference  frames  $I_{\tau}$. The quantities with an index without parenthesis, for the field components in the laboratory system, are described in (\ref{eq:mff1}), with $t_{i}= \gamma \tau_{i} $.
The quantities $\beta$ and $\gamma$ do not do not depend on $\tau$ and are not supplied by index i.\\
\indent It is easy to see from (\ref{eq:field_Instantaneous}) and (\ref{eq: alpha}) that in terms of $\tau$ all these fields are periodic with the 
period $T_{\gamma}= 2 \pi/ \Omega_{\gamma}$, with $\Omega_{\gamma}=\Omega_0 \gamma$. 
\section{ Calculation of the Force $ <F^{(i)}> \equiv < \frac{2}{3}(\dot{\vec{p}} \times \vec{H}_{0})^{(i)}>$}
\subsection{Rate of Change  of the Dipole of the z-Oscillator in Zero-Point Field.} 
To calculate the rate of change of the z-oscillator dipole, $\dot{p}^{(z)}(\tau) \equiv e \, v^{(z)}(\tau) = e \frac{dz}{d \tau}$,  we use the motion equation of a z-oscillator in the coordinate system   defined  as a set of all instantaneous inertial frames $I_{\tau}$ . It is taken in the form similar to (14) in \cite{boyer1984} for hyperbolic motion and justified exactly the same way:
\begin{eqnarray}
\label{eq:zOscillator}
m \frac{d^2 z}{ d \tau^2}=-m\omega^2_0 z + \frac{2}{3} \frac{e^2}{c^3}[ \frac{d^3z}{d \tau^3} - \frac{a^2}{c^2}\frac{dz}{d \tau} ] + e E^{(z)}_0(0, \tau), \;\;\; z \equiv x^{(3)}
\end{eqnarray}
The only  difference between them is that
the constant acceleration $ \vec{a}$ in this equation, unlike the hyperbolic motion,  is centripetal and directed against axis $x^{(1)}$
 because  4-vector acceleration of the observer in the associated  instantaneous reference frame $I_{\tau}$ is constant in both magnitude and direction 
\begin{eqnarray}
\dot{U}_{(a)}= \mu^i_{(a)}\dot{U}_i=
(-R \Omega_0^2 \gamma^2,0,0,0).
\end{eqnarray} 
Solution to (\ref{eq:zOscillator}) is
\begin{eqnarray}
\label{eq:dipoleChange}
\dot{p}^{(3)}(\tau) =
\label{eq:I_one}
\sum_{n=-\infty}^{\infty} \frac{e^2}{m}\frac{(i \omega_n) \exp(i \omega_n \tau)}{\omega_0^2 - \omega_n^2 +i \Gamma(\omega_n^3 + \omega_n \frac{a^2}{c^2})} \frac{1}{T_{\gamma}} \int_{-T_{\gamma}/2}^{T_{\gamma}/2}d\tau_1 E^{(3)}_0(\tau_1) \exp(-i \omega_n \tau_1),   \nonumber \\
 \omega_n=\frac{2 \pi}{T_{\gamma}} n, \; \; n=0, \pm 1, \pm 2,..., \; \; T_{\gamma}=\frac{T}{\gamma}, \;\;\;
\Gamma=\frac{2}{3} \frac{e^2}{m c^3}.
\end{eqnarray}
It is a rate of change  of the dipole of the rotating z-oscillator induced by zero-point field and which is used in (\ref{eq:FordForce}).
\subsection{The Force $ <F^{(x)}>$  in Terms of the Correlation Function and the Oscillator Selectivity Function.}
The (x)-component of the component of the force  ($\ref{eq:FordForce}$), which is due to the oscillator dipole change, in an instantaneous inertial reference frame $I_{\tau}$ after averaging is 
\begin{eqnarray}
< \tilde{F}^{(x)} >  =\frac{2}{3} < (\dot{\vec{p}}(\tau) \times \vec{H}_0(\tau))^{(x)} >,
\end{eqnarray}
where the symbol $< ... >$  means averaging over random phases and is discussed in the next subsections below.\\
\indent Using  (\ref{eq:dipoleChange})  this expression can be written in a more transparent  and convenient for calculation and interpretation form :
\begin{eqnarray}
\label{eq:X_Component_Of_Force}
<\tilde{F}^{(x)}> = -\frac{2}{3}\frac{e^2}{m c} \frac{1}{T_{\gamma}} \int_{-T_{\gamma}/2}^{T_{\gamma}/2} d(\tau_1-\tau_2) f_d(\tau_1-\tau_2) <E^{(z)}(\tau_1) H^{(y)}(\tau_2)>_d, 
\end{eqnarray}
where 
\begin{eqnarray}
\label{eq:I^{zy}_dd}
<E^{(z)}(\tau_1)H^{(y)}(\tau_2)>_d=\frac{\hbar c k^4_0 }{2
\pi^2}\int do [-\hat{k}_x\gamma \cos
\frac{\delta}{2}-\hat{k}_y\gamma
\sin\frac{\delta}{2}+\hat{k}_x\hat{k}_y \beta \gamma
+\frac{1}{2}\beta \gamma \sin\delta (1+ \hat{k}_z^2)] \times
 \nonumber \\
  \;\;\sum_{n=0}^{\infty} n^3 \cos n F
\end{eqnarray}
 is the correlation function, Appendix \ref{section: Calculaion of the CF}, with
 \begin{eqnarray}
  F \equiv \delta [1-\beta \frac{\sin \delta /2}{\delta/2}\sin \theta \sin
  \phi], \;\;\;\; k_0 =\Omega_0/c, \;\;\; \delta=\Omega_0 \gamma (\tau_2-\tau_1),
 \end{eqnarray}
and
\begin{eqnarray}
\label{eq:f_d}
f_d(\tau_1-\tau_2) \equiv \sum_{n=-\infty}^{\infty} \frac{(i \omega_n) \exp(-i \omega_n(\tau_1- \tau_2))}{\omega_0^2 - \omega_n^2 +i \Gamma(\omega_n^3 + \omega_n \frac{a^2}{c^2})}
\end{eqnarray}
is the oscillator selectivity function depending on inner structure of the oscillator, Section  \ref{section: Calculaion of the CF}. \\
( The symbol F without indexes is just a designation, not a force!) \\
\indent We see  that  both the correlation function and the selectivity function  depend on
 difference of times, $\tau_1-\tau_2$, and then the force does not depend on time. Therefore we have omitted time in the function $<\tilde{F}^{(x)}>$ in  (\ref{eq:X_Component_Of_Force}).  Index "d" means that corresponding quantities contain discrete variable $\omega_n$. \\
 \indent\label{sec:Analysis}
 The term $ \sum_{n=0}^{\infty}$ in the correlation function, with the help of Abel-Plana formula
( \cite{Bateman1953},\cite{grib1980}, \cite{MT1996}, \cite{Evgrafov1968})
  \begin{eqnarray}
  \label{eq:AbelPlana}
   \sum_{n=0}^\infty \, f(n)= \int_0^\infty
  f(x)\,dx + \frac{f(0)}{2} +i \,\int_0^\infty \,dt \,
  \frac{f(it)-f(-it)}{e^{2 \pi t}-1},
  \end{eqnarray}
  can be  represented as a sum of two functions:
  \begin{eqnarray}
  \label{eq:sum}
  S_d \equiv \sum_{n=0}^{\infty}n^3 \cos nF = 
    \frac{1}{\Omega_0^4} \int _0^{\infty} d \, \omega \omega^3 \cos( \omega \tilde{F})
   + \frac{1}{\Omega_0^4}\int_0^{\infty} d\omega \frac{2 \omega^3 \cosh(\omega \tilde{F})}{e^ {2\pi \omega/\Omega_0}-1} \equiv S_c + S_r , 
  \end{eqnarray}
  with $\tilde{F}=\frac{F}{\Omega_0}$. \\
The integrals in this expression can be computed: 
\begin{eqnarray}
S_c= \frac{6}{F^4}, \;\; S_r= [\frac{3 -2 \sin^2(F/2)}{8
\sin^4(F/2)} -\frac{6}{F^4}]
\end{eqnarray}
and $S_d$ and $S_c$ are divergent but their difference $S_r= S_d -S_c $ converges at
  $|\tau_2-\tau_1|\rightarrow 0$:
  \begin{eqnarray}
  \label{(eq:S_renormalized)}
  \lim_{F\rightarrow 0}S_c = \frac{6}{F^4}= \infty, \;\; \lim_{F\rightarrow 0}S_d =
  \infty, \;\;\;
  \lim_{F \rightarrow 0}S_r=\frac{1}{120}
  \end{eqnarray}
  For further interpretation a presentation of $S_d$ as a series of fractions can also be useful:  
  \begin{eqnarray}
  \label{eq:I_two}
   S_d= \frac{6}{F^4} +
  6  \sum_{n=1}^{\infty} \frac{1}{(2\pi n)^4} [
  \frac{1}{(1+\frac{F}{2\pi n})^4}  + \frac{1}{(1-\frac{F}{2\pi
  n})^4} ] = 6  \sum_{n= -\infty}^{\infty} 
    \frac{1}{(2 \pi n + F )^4}.
  \end{eqnarray}
  So the correlation function with a discrete spectrum is a sum of two parts:
\begin{eqnarray}
<E^{(z)}(\tau_1)H^{(y)}(\tau_2)>_d= <E^{(z)}(\tau_1)H^{(y)}(\tau_2)>_c +<E^{(z)}(\tau_1)H^{(y)}(\tau_2)>_r
\end{eqnarray}  
where the first one, with index c,  corresponds to $S_c$ with continuous spectrum and the second one to $S_r$. Then subtracting $<...>_c$ correlation function from $<...>_d$ we obtain convergent correlation function describing net contribution of the periodicity  
\begin{eqnarray}
\label{eq:I^{zy}_d}
<E^{(z)}(\tau_1)H^{(y)}(\tau_2)>_r=\frac{\hbar c k^4_0}{2
\pi^2}\int do [-\hat{k}_x\gamma \cos
\frac{\delta}{2}-\hat{k}_y\gamma
\sin\frac{\delta}{2}+\hat{k}_x\hat{k}_y \beta \gamma
+\frac{1}{2}\beta \gamma \sin\delta (1+ \hat{k}_z^2)] \times
 \nonumber \\
  \frac{1}{\Omega_0^4}\int_0^{\infty} d\omega \frac{2 \omega^3 \cosh(\omega \tilde{F})}{e^ {2\pi \omega/\Omega_0}-1}
\end{eqnarray}
or after some simplifications, Appendix \ref{sec:SimplificationOfRenCorFunct},
 \begin{eqnarray}
\label{eq:RenormCorFunc}
  <E^{(z)}(\tau_1)H^{(y)}(\tau_2)>_r=\frac{\hbar c k^4_0}{2
  \pi^2} \times \frac{1}{2} \beta \gamma \sin \delta   \int_0^{\infty} dx \frac{2 x^3 \cosh(x
  \delta) )}{e^ {2\pi x}-1} \int_{0}^{\pi}d \theta \sin \theta (1 + \cos^2 \theta ) \times \nonumber \\
   \int_{0}^{2 \pi}
  d \phi \cosh(2 x \beta \sin (\delta/2 ) \sin \theta \sin \phi)  
  \end{eqnarray}  
\indent  Modification from 
$<E^{(z)}(\tau_1)H^{(y)}(\tau_2)>_d$ to $<E^{(z)}(\tau_1)H^{(y)}(\tau_2)>_r$ is similar to a local renormalization procedure used in Casimir effect theory \cite{MT1996}(1.31). 
The renormalized correlation function $<E^{(z)}(\tau_1)H^{(y)}(\tau_2)>_r$ corresponds to a net change in vacuum fluctuations
 spectrum, which is due to the periodicity condition, and it is reasonable to use it instead of $<E^{(z)}(\tau_1)H^{(y)}(\tau_2)>_d$ in calculation of the observed force. 
  \subsection{The Expression for the Renormalized Force $<F^{(x)}>_r \equiv <\frac{2}{3}(\dot{\vec{p}} \times \vec{B}_{0})^{(x)}>_r$. The case $\beta \ll 1$, $\omega_0 \ll \Omega_0$.}
The exact expression for the force $ <\frac{2}{3}(\dot{\vec{p}} \times \vec{B}_{0})^{(i)}>_r $ corresponding to
the renormalized correlation function 
is determined by  three equations:  (\ref{eq:X_Component_Of_Force}),  (\ref{eq:f_d}), and (\ref{eq:RenormCorFunc}).
  To represent it in simpler form turned out to be difficult. But in the case $\omega_0 \ll \Omega_0$
  the selectivity function $f_d$ reduces to (\ref{f_d})  and the force becomes  
 \begin{eqnarray}
 <\frac{2}{3}(\dot{\vec{p}} \times \vec{B}_{0})^{(x)}>_r =-\frac{1}{3 \pi^3}\;\frac{e^2 \hbar}{m c^4}\;\Omega_0^3 \beta \sum_{n=1}^{\infty} 
 \frac{1}{n \; [1 +(\Gamma \Omega_0 \gamma)^2 n^2(1 + \frac{\beta^2 \gamma^2}{n^2})^2]} \; \int_{0}^{\pi} d \theta
 \sin \theta (1 + \cos^2 \theta) \int_{0}^{2 \pi} d\phi \times \nonumber \\
  \int_{-\pi}^{\pi} d\delta \sin(n\delta) \sin\delta \;
 \int_{0}^{\infty}dx \frac{x^3\;\cosh(x \delta)\;\cosh(2 x \beta \sin(\delta/2) \sin\theta \sin\phi)}{exp(2 \pi x)-1}
 , \;\;\; \omega_0 \ll \Omega_0.
 \end{eqnarray} 
  This expression can also be significantly simplified  when $\beta \ll 1$ and $\gamma \approx 1$. In the first order of $\beta$ the force is 
 \begin{eqnarray}
 <F^{(x)}>_r \approx <F^{(x)}>_r \mid_{(\beta=0)} + \frac{\partial <F^{(x)}>_r }{\partial \beta} \mid_{(\beta=0)} \times \beta.
 \end{eqnarray}
 In this case $\cosh(2 x \beta \sin(\delta/2) \sin\theta \sin\phi) \mid_{(\beta=0)} = 1$, integration over  $\phi$, $\theta$ and $\delta$  can be separated and easily carried out. Then
 \begin{eqnarray}
 <F^{(x)}>_r =-\frac{32}{9 \pi^2} \frac{e^2 \hbar}{m c^4} \Omega_0^3 \beta \;\sum_{n=1}^{\infty} \frac{(-1)^{n+1}}{1 +
 (\Gamma \Omega_0)^2 n^2} \; \int_{0}^{\infty}dx \frac{x^4}{exp(\pi x)[x^2 +(n+1)^2][x^2+(n-1)^2]}, \nonumber \\ \omega_0 \ll \Omega_0, \;\; \beta \ll 1, \;\; \gamma \approx 1.
 \end{eqnarray} 
 Retaining only the first term in the sum and ignoring all other fast decreasing terms we have
 \begin{eqnarray}
 \label{eq:force_first_component}
 <F^{(x)}>_r =-(0.013) \frac{32}{9 \pi^2} \frac{e^2 \hbar}{m c^4}   \; \frac{\Omega_0^3}{1 +
 (\Gamma \Omega_0)^2 }\;\beta
 \end{eqnarray}
 or
 \begin{eqnarray}
  \label{eq:force_first_component}
  <F^{(x)}>_r =-(0.013) \frac{32}{9 \pi^2} \frac{e^2 \hbar}{m c^5}   \; \frac{\Omega_0^4}{1 +
  (\Gamma \Omega_0)^2 }\;R
  \end{eqnarray}
  \indent Two approximations need to be explained. First, a small speed,
  $\beta \ll 1, $ does not prevent large angular velocity. It just means that rotation radius can be very small. The condition $\omega_0 \ll \Omega_0$
  means that instead of the oscillator we deal with a point-like particle
  but the particle oscillation is still restricted to only z-direction. \\
  \indent So the centripetal renormalized force $<F^{(x)}>_r$, at least  in non relativistic case,  is directed to the center of rotation and proportional to the radius
of the circle. 
 \section{Calculation of the Force  $ <F_1^{(1)}> \equiv <\frac{1}{3}p_{(3)}\partial^{(1)} E^{(3)}>$.}
\subsection{The Force  $<F_1^{(1)}>$ in Terms of a Correlation Function and a Selectivity Function.}
 For the x-component of the force acting on the z-oscillator this expression gets the form 
\begin{eqnarray} 
\label{eq:forceF_1}
<F_1^{(1)}> \equiv<\frac{1}{3}p_{(3)}\partial^{(1)} E^{(3)}> = 
\sum_{n=-\infty}^{\infty} \frac{e^2}{m}\frac{1}{\omega_0^2 - \omega_n^2 +i \Gamma(\omega_n^3 + \omega_n \frac{a^2}{c^2})} \frac{1}{3}\frac{1}{T_{\gamma}} \int_{-T_{\gamma}/2}^{T_{\gamma}/2}d\tau_1  \exp(-i \omega_n (\tau_1-\tau_2))\times \nonumber \\ <E^{(3)}(\tau_1)\partial^{(1)}E^{(3)}(\tau_2)>, \nonumber \\
\end{eqnarray}
or after changing the order of integration and summation 
\begin{eqnarray}
\label{eq:forceF_11} 
<F_1^{(1)}>= 
 \frac{e^2}{3 m} \; \frac{1}{T_{\gamma}}\int_{-T_{\gamma}/2}^{T_{\gamma}/2}d(\tau_1 - \tau_2) f_{1,d}(\tau_1-\tau_2) <E^{(3)}(\tau_1)\partial^{(1)}E^{(3)}(\tau_2)>.  
\end{eqnarray}
The function
\begin{eqnarray}
 \label{eq:f_{1,d}}
f_{1,d}(\tau_1-\tau_2)=-2 \sum_{n=1}^{\infty}\frac{1}{\omega_n^2}
\frac{ \sin[\omega_n(\tau_1-\tau_2) + \phi_n]}
{\sqrt{(1 - \frac{\omega_0^2}{\omega_n^2})^2 + \Gamma^2 \omega_n^2(1+ \frac{a^2}{
\omega_n^2 c^2})^2}}, 
\end{eqnarray}
is similar to the function $f_d$ in (\ref{eq:f_d}). \\
The angle $\phi_n$ in the last expression is defined by the equation
 \begin{eqnarray}
\sin \phi_n =\frac{1-\frac{\omega_0^2}{\omega_n^2}}{\sqrt{(1 - \frac{\omega_0^2}{\omega_n^2})^2 + \Gamma^2 \omega_n^2(1+ \frac{a^2}{
\omega_n^2 c^2})^2}}, 
\end{eqnarray}
\subsection{Calculation of the Correlation Function $<E^{(3)}(\tau_1)\partial^{(1)}E^{(3)}(\tau_2)>$. }
We do not supply $F_1^{(1)}$ with time arguments because we will see later that it does not depend on time. \\
Introducing local coordinates $x^{(\nu)}(\tau)$ in an instantaneous inertial reference frame, defined by $\mu^i_{(\nu)}(\tau)$,
connected with the center of the rotating z-oscillator,
\cite{moller}, \cite{moller72}(9.88)
\begin{eqnarray}
\label{eq:LocalCoordinates}
x^i(\tau)=x_0^i(\tau) + \mu^i_{(\nu)}(\tau)x^{(\nu)}, \; \; \; i, \nu =1,2,3,4
\end{eqnarray}
and using the formulas 
\begin{eqnarray}
E^{(3)}=-E_{(3)} =- F_{(34)}=-\frac{\partial x^i}{\partial x^{(3)}} \frac{\partial x^k}{\partial x^{(4)}} F_{ik}, \nonumber
\\
\frac{\partial E^{(3)}}{\partial x^{(1)}}=-\frac{\partial x^i}{\partial x^{(3)}}\frac{\partial x^k}{\partial x^{(4)}}
\frac{\partial x^l}{\partial x^{(1)}} \frac{\partial F_{ik}}{\partial x^{(l)}}, \; \; \;\;
\frac{\partial}{\partial x^{(1)}}(\frac{\partial x^i}{\partial x^{(3)}}\frac{\partial x^k}{\partial x^{(4)}} ) =0,
\end{eqnarray}   
we can easily  represent the correlation function in terms of zero-point fields in the laboratory system as
\begin{eqnarray}
<E^{(3)}(\tau_1) \frac{\partial E^{(3)}(\tau_2)}{\partial x^{(1)}} > = 
<[ \beta \gamma \sin \alpha_1 H_2(\tau_1) + \beta \gamma \cos \alpha_1 H_1(\tau_1) -\gamma E_3(\tau_1)] 
\times 
[ \frac{\beta \gamma}{2} 
\sin(2 \alpha_2) \frac{\partial H_2(\tau_2)}{\partial x^1}
+ \nonumber \\
\beta \gamma \cos^2 \alpha_2 \frac{\partial H_1(\tau_2)}{\partial x^1}- 
\gamma \cos \alpha_2 \frac{\partial E_3(\tau_2)}{\partial x^1} + 
\beta \gamma \sin^2 \alpha_2 \frac{\partial H_2(\tau_2)}{\partial x^2} +
\frac{\beta \gamma}{2} \sin(2 \alpha_2)\frac{\partial H_1(\tau_2)}{\partial x^2}- \gamma \sin \alpha_2 
\frac{\partial E_3(\tau_2)}{\partial x^2}]>.
\nonumber \\
\end{eqnarray}
Spatial derivatives of the fields are taken at  locations $\vec{x}_0(t_i)$ of the oscillator center(\ref{eq:OscillatorRotation}) at
the times $t_1$ and $t_2$. So from
(\ref{eq:ff1})  we have
\begin{eqnarray}
\frac{\partial \vec{E}(\vec{x}, t_i)}{\partial x^l}\vert_{\vec{x}=\vec{x}_0(t_i)} = k_0 \sum_{n=0}^{\infty} \sum_{\lambda=1}^2 \int d\theta d \phi \sin \theta 
(-k_n^l) k_n^2 \hat{\epsilon}(\vec{k}_n, \lambda) h_0(\omega_n) \sin[\vec{k}_n \vec{x}_0(t_i) - \omega_n t_i -\theta(\vec{k_n}, \lambda)]
\nonumber \\
\frac{\partial \vec{H}(\vec{x}, t_i)}{\partial x^l}\vert_{\vec{x}=\vec{x}_0(t_i)} = k_0 \sum_{n=0}^{\infty} \sum_{\lambda=1}^2 \int d\theta d \phi \sin \theta 
(-k_n^l) k_n^2 [\hat{k},\hat{\epsilon}(\vec{k}_n, \lambda)] h_0(\omega_n) \sin[\vec{k}_n \vec{x}_0(t_i) - \omega_n t_i -\theta(\vec{k_n}, \lambda)]
\nonumber \\
\end{eqnarray} 
\indent After simple but long and tedious calculations, similar to ones used for (\ref{eq:I^{zy}_dd}), we come to the expression 
for the correlation function $<E^{(3)}(\tau_1)\partial^{(1)}E^{(3)}(\tau_2)>$: 
\begin{eqnarray}
\label{eq:CorFunctionE_3D_1E^3} 
<E^{(3)}(\tau_1)\partial^{(1)}E^{(3)}(\tau_2)>=-\frac{\hbar c}{2 \pi^2} k_0^5 \int d\theta d \phi \sin \theta \times 
\{
\hat{k}_1 \cos \frac{\delta}{2}((\beta \gamma)^2 \cos \delta + \gamma^2) + \hat{k}_2 \sin\frac{\delta}{2}((\beta \gamma)^2 \cos \delta + \nonumber \\
\gamma^2) + \hat{k}_1\hat{k}_2 (-2)\beta \gamma^2 \cos^2 \frac{\delta}{2} + \hat{k}_2^2(-\beta \gamma
^2) \sin\delta + \hat{k}_1^3(\beta \gamma)^2 (-\cos^3 \frac{\delta}{2}) +
 \hat{k}_1^2 \hat{k}_2(\beta \gamma)^2(-\sin \frac{\delta}{2} \cos^2 \frac{\delta}{2})+ \nonumber \\
 \hat{k}_1 \hat{k}_2^2(\beta \gamma
 )^2 \cos \frac{\delta}{2}\sin^2\frac{\delta}{2} +\hat{k_2^3}(\beta \gamma)^2 \sin^3\frac{\delta}{2}+
 \hat{k}_1\hat{k}_3^2(-\gamma^2)\cos\frac{\delta}{2}+ \hat{k}_2\hat{k}_3^2(-\gamma^2)\sin\frac{\delta}{2} 
 \} \times  \sum_{n=0}^{\infty}n^4 \sin (-n F)\nonumber \\
\end{eqnarray} 
As in (\ref{eq:I^{zy}_d}),  the  correlation function depends on difference of times $\delta=\alpha_2 -\alpha_1=\Omega_0 \gamma (\tau_2 -\tau_1)$ as it is supposed to be for the correlation function  of a stationary process of rotation.
\subsection{Analysis and Renormalization of the Correlation Function $<E^{(3)}(\tau_1)\partial^{(1)}E^{(3)}(\tau_2)>$.}
Using the Abel-Plana formula again we obtain \\
\begin{eqnarray}
S_1 \equiv \sum_{n=0}^{\infty} n^4 \sin (+nF)= S_{1,c} + S_{1,r}, \\
S_{1,c}= \int_{0}^{\infty}dt \; t^4 \sin tF, \\
\label{eq:S_1S_2}
S_{1,r}= - \int_{0}^{\infty} dt \frac{2 t^4 \sinh tF}{\exp(2\pi t) -1}.
\end{eqnarray}
The first integral is divergent if $| \tau_2 -\tau_1| \rightarrow 0$ ( and  $F \rightarrow 0 $) because
\begin{eqnarray}
S_{1,c} =  \frac{24}{F^5}.
\end{eqnarray}
The second integral can also be represented as a sum
\begin{eqnarray}
S_{1,r}= 24\; \sum_{n=1}^{\infty} [\frac{1}{ (2 \pi n + F)^5} - \frac{1}{(2 \pi n -F)^5}],
\end{eqnarray}
and then the correlation function $S_1$ can also be given in the form
\begin{eqnarray}
S_1 = 24 \sum_{-\infty}^{\infty} \frac{1}{(2 \pi n +F)^5}.
\end{eqnarray}
It is obvious that $S_{1,r}$ is convergent and  for $F \ll 1$, 
\begin{eqnarray}
S_{1,r} \approx - \frac{15}{16 \times 945} F.
\end{eqnarray}
For the reasons, explained in Section \ref{sec:Analysis}, the renormalized correlation function corresponding $S_{1,r}$ will be used to calculate the renormalized force $F_{1,r}^{(1)}$. It has the form:
\begin{eqnarray}
\label{eq:CorFunctionE_3D_1E^3Ren} 
<E^{(3)}(\tau_1)\partial^{(1)}E^{(3)}(\tau_2)>_r=-\frac{\hbar c}{2 \pi^2} k_0^5 \int d\theta d \phi \sin \theta \times 
\{
\hat{k}_1 \cos \frac{\delta}{2}((\beta \gamma)^2 \cos \delta + \gamma^2) + \hat{k}_2 \sin\frac{\delta}{2}((\beta \gamma)^2 \cos \delta + \nonumber \\
\gamma^2) + \hat{k}_1\hat{k}_2 (-2)\beta \gamma^2 \cos^2 \frac{\delta}{2} + \hat{k}_2^2(-\beta \gamma
^2) \sin\delta + \hat{k}_1^3(\beta \gamma)^2 (-\cos^3 \frac{\delta}{2}) +
 \hat{k}_1^2 \hat{k}_2(\beta \gamma)^2(-\sin \frac{\delta}{2} \cos^2 \frac{\delta}{2})+ \nonumber \\
 \hat{k}_1 \hat{k}_2^2(\beta \gamma
 )^2 \cos \frac{\delta}{2}\sin^2\frac{\delta}{2} +\hat{k_2^3}(\beta \gamma)^2 \sin^3\frac{\delta}{2}+
 \hat{k}_1\hat{k}_3^2(-\gamma^2)\cos\frac{\delta}{2}+ \hat{k}_2\hat{k}_3^2(-\gamma^2)\sin\frac{\delta}{2} 
 \} \times 
 (+1)\int_{0}^{\infty} dt \frac{2 t^4 \sinh tF}{\exp(2\pi t) -1}. \nonumber \\  
\end{eqnarray} 
\subsection{The Expression for the Force $<F_{1}^{(1)}>_r \equiv <\frac{1}{3}p_{(3)}\partial^{(1)} E^{(3)}>_r$. The Case $\beta \ll 1$.}
The expression (\ref{eq:CorFunctionE_3D_1E^3Ren}), pretty complicated,can be significantly simplified if $\beta \ll
1$. Then $F \approx \delta$, the last integral over t does not depend on $\theta$ and $\phi$, and  the integral over 
angular variables $\theta$ and $\phi$ can be easily found. The only non zero contribution comes from the $\hat{k}_2^2$  term.\\
 The renormalized correlation function takes the form:
\begin{eqnarray}
\label{eq:CorFunctionE_3D_1E^3RenSimpl} 
<E^{(3)}(\tau_1)\partial^{(1)}E^{(3)}(\tau_2)>_r=
 (-\frac{\hbar c}{2 \pi^2} k_0^5) \frac{4 \pi}{3}(-\beta \gamma^2) \sin \delta  \times  \int_{0}^{\infty} dt \frac{2 t^4 \sinh t \delta}{\exp(2\pi t) -1}, \;\;\; \beta \ll 1.  
\end{eqnarray} 
With this renormalized correlation function and $f_{1,d}$ defined in  (\ref{eq:f_{1,d}}) , the renormalized force  $<F_{1}^{(1)}>_r$ corresponding to (\ref{eq:forceF_11} ) becomes
\begin{eqnarray} 
<F_{1}^{(1)}>_r=
 \frac{e^2}{3 m} \; \frac{1}{T_{\gamma}}\int_{-T_{\gamma}/2}^{T_{\gamma}/2}d(\tau_1 - \tau_2) \times
  (-2) \sum_{n=1}^{\infty} \frac{1}{\omega_n^2} \frac{\sin[\omega
  _n(\tau_1 - \tau_2) + \phi_n]}{\sqrt{(1 - \frac{\omega_0^2}{\omega_n^2})^2 + \Gamma^2 \omega_n^2(1+ \frac{a^2}{
  \omega_n^2 c^2})^2}} \times \nonumber \\
 (-\frac{\hbar c}{2 \pi^2} k_0^5) \frac{4 \pi}{3}(-\beta \gamma^2) \sin \delta  \times 
 (+1) \int_{0}^{\infty} dx \frac{2 x^4 \sinh x \delta}{\exp(2\pi x) -1},  \;\;\; \beta \ll 1, \;\;\;\; \delta =\Omega_0 \gamma (\tau_2 - \tau_1). 
 \nonumber \\  
\end{eqnarray}
Moving all time-dependent terms to the right we obtain
\begin{eqnarray} 
<F_{1}^{(1)}>_r=
 \frac{e^2}{3 m} \;(-2) \;  (-\frac{\hbar c}{2 \pi^2} k_0^5) \frac{4 \pi}{3}(-\beta \gamma^2) \; (+1)\;\times
  \sum_{n=1}^{\infty} \frac{1}{\omega_n^2} \frac{1 }{\sqrt{(1 - \frac{\omega_0^2}{\omega_n^2})^2 + \Gamma^2 \omega_n^2(1+ \frac{a^2}{
  \omega_n^2 c^2})^2}} \times 
 \int_{0}^{\infty} dx \frac{2 x^4 \;I_{1,n}(x) }{\exp(2\pi x) -1}, \nonumber \\
 \end{eqnarray}
 where
 \begin{eqnarray}
   I_{1,n}(x) \equiv \frac{1}{T_{\gamma}}\int_{-T_{\gamma}/2}^{T_{\gamma}/2}d(\tau_1 - \tau_2) \;\sin[\omega
   _n(\tau_1 - \tau_2) + \phi_n]\; \sin \delta  \; \sinh x \delta  = \nonumber \\
   \frac{1}{\pi}(-1)^{n+1}\sinh (\pi x) \sin\phi_n
 \frac{n^2-1-x^2}{[(n-1)^2 + x^2][(n+1)^2 + x^2]}, 
\end{eqnarray}
or, after some rearrangements,
\begin{eqnarray} 
<F_{1}^{(1)}>_r= -\frac{4}{9 \pi^2} \frac{e^2\hbar c}{ m} k_0^5\beta \gamma^2 \times
  \sum_{n=1}^{\infty} \frac{1}{\omega_n^2} \frac{(-1)^{n+1} \sin \phi_n \; J_{1,n}}{\sqrt{(1 -\frac{\omega_0^2}{\omega_n^2})^2 + \Gamma^2 \omega_n^2(1+ \frac{a^2}{
  \omega_n^2 c^2})^2}},
  \end{eqnarray}
  and 
  \begin{eqnarray}
  \label{eq:F_1^{(1)}}
<F_{1}^{(1)}>_r=  -\frac{4}{9 \pi^2} \frac{e^2\hbar }{ m c^4} \Omega_0^3 \beta \times
    \sum_{n=1}^{\infty}
     \frac{(-1)^{n+1}\;(1 -\frac{\omega_0^2}{\Omega_0^2 n^2}) \; J_{1,n}}{n^2 ((1-\frac{\omega_0^2}{\Omega_0^2 n^2})^2 + \Gamma^2 \Omega_0^2 n^2)}, \;\; \beta \ll 1, \;\; \gamma \approx 1, 
  \\ 
J_{1,n} \equiv \int_{0}^{\infty} dx \frac{ x^4 \exp (-\pi x) (n^2 -1 -x^2) }{[(n-1)^2 + x^2][(n+1)^2 + x^2]}. \nonumber \\
 \end{eqnarray}
\section{Calculation of the Force $<F_2^{(1)}> \equiv <\frac{2}{3} p_{(3)} \partial^{(3)} E_0^{(1)}> $}
\subsection{The Force $F^{(1)}_2$ in Terms of a Correlation Function and a Selectivity Function. }
For the x-component of the  $F^{(1)}_2$ the expression is \\
\begin{eqnarray}
\label{eq:forceF_21} 
<F_2^{(1)}>= 
 \frac{2 e^2}{3 m} \; \frac{1}{T_{\gamma}}\int_{-T_{\gamma}/2}^{T_{\gamma}/2}d(\tau_1 - \tau_2) f_{1,d}(\tau_1-\tau_2) <E^{(3)}(\tau_1)\partial^{(3)}E^{(1)}(\tau_2)>.  
\end{eqnarray}
The  function $f_{1,d}$ is given in ( \ref{eq:f_{1,d}} ), and the correlation function $<E^{(3)}(\tau_1) \partial^{(3)}
E^{(1)}(\tau_2)>$ is \\
\begin{eqnarray}
<E^{(3)}(\tau_1) \partial^{(3)}E^{(1)}(\tau_2)> =< [- \gamma E_3(\tau_1) + \beta \gamma \cos(\alpha_1)H_1(\tau_1) + 
\beta \gamma \sin(\alpha_1) H_2(\tau_1) 
]  \times  
 [ - \beta \gamma \frac{\partial H_3(\tau_2)}{\partial x^3} - \nonumber \\
 \gamma \cos(\alpha_2) 
\frac{\partial E_1(\tau_2)}{\partial x^3} -\gamma \sin(\alpha_2) \frac{\partial E_2(\tau_2)}{ \partial x^3}] >.
\end{eqnarray}
\subsection{ The Correlation Function $<E^{(3)}(\tau_1) \partial^{(3)}E^{(1)}(\tau_2)>$ and its Renormalization.}
After tedious calculations, similar to ones leading to ( \ref{eq:I^{zy}_dd} ) and described in Appendix 
\ref{section: Calculaion of the CF}, we come to the expression 
\begin{eqnarray}
<E^{(3)}(\tau_1) \partial^{(3)}E^{(1)}(\tau_2)> =-\frac{\hbar c}{2 \pi^2}k_0^5 \gamma^2 \int do \; \hat{k}_3^2 \; [(1-
 \beta^2) \hat{k_1} \cos(\delta/2)+ (1+\beta^2)\hat{k}_2 \sin(\delta/2)] \times \sum_{0}^{\infty} n^4 \sin nF
\end{eqnarray}
with the sum term like one in  (\ref{eq:I^{zy}_dd}). Then the renormalized correlation function is  
\begin{eqnarray}
<E^{(3)}(\tau_1) \partial^{(3)}E^{(1)}(\tau_2)>_r =-\frac{\hbar c}{2 \pi^2}k_0^5 \gamma^2 \int do \; \hat{k}_3^2 \; [(1-
 \beta^2) \hat{k_1} \cos(\delta/2)+ (1+\beta^2)\hat{k}_2 \sin(\delta/2)] \times\nonumber \\
  (-1) \int_{0}^{\infty} dt
 \frac{2 t^4 \sinh (tF)}{\exp(2 \pi t) -1}.
\end{eqnarray}
\subsection{The Force $<F_2^{(1)}> \equiv <\frac{2}{3} p_{(3)} \partial^{(3)} E_0^{(1)}> $ for $\beta \ll 1$.}
It is easy to see that
\begin{eqnarray}
<E^{(3)}(\tau_1) \partial^{(3)}E^{(1)}(\tau_2)>_{r, \; \beta \ll 1, \; \gamma \approx 1}= \frac{\hbar c}{2 \pi^2} k_0^5
\int d0 \; \hat{k}_3\; [\hat{k}_1 \cos(\delta/2) + \hat{k}_2 \sin(\delta/2)] \times \nonumber \\ \int_{0}^{\infty}
\;dt \; \frac{
2 t^4 \{ \sinh(t \delta) -  2 \; \hat{k}_2 \; t \; \beta \sin(\delta/2) \cosh(t \delta) \} }
{\exp(2 \pi t) -1},      
\end{eqnarray}
and, because
\begin{eqnarray}
\int\;d0 \hat{k}_3 \hat{k}_1= \int\;d0 \hat{k}_3 \hat{k}_2= \int\;d0 \hat{k}_3 \hat{k}_1 \hat{k}_2= \int\;d0 \hat{k}_3 \hat{k}_2^2= 0, 
\end{eqnarray}
this renormalized correlation function is zero,
\begin{eqnarray}
<E^{(3)}(\tau_1) \partial^{(3)}E^{(1)}(\tau_2)>_{r, \; \beta \ll 1, \; \gamma \approx 1}= 0,
\end{eqnarray}
and the corresponding (renormalized ) force is also zero:
\begin{eqnarray}
\label{eq:F_2^{(1)}}
<F_2^{(1)}>_{r, \; \beta \ll 1, \; \gamma \approx 1} \equiv <\frac{2}{3} p_{(3)} \partial^{(3)} E_0^{(1)}>_{r, \; \beta \ll 1, \; \gamma \approx 1} =0.
\end{eqnarray}
 \section{The Total Renormalized Force $<\tilde{F}^{(1)}>_r$: Estimation for $\beta \ll 1$, $\gamma \approx 1$, and $\omega_0 \ll \Omega_0$.}
From ($\ref{eq:force_first_component}$), ($\ref{eq:F_1^{(1)}}$) and ($\ref{eq:F_2^{(1)}}$) we obtain the expression for the total renormalized force:
  \begin{eqnarray}
<F^{(1)}>_r + <F_{1}^{(1)}>_r + <F_{2}^{(1)}>_r = 
 <\frac{2}{3}(\dot{\vec{p}}\times\vec{H}_{0})^{(1)}>_r + <\frac{1}{3}p_{(3)}\partial^{(1)} E^{(3)}>_r+
 <\frac{2}{3} p_{(3)} \partial^{(3)} E_0^{(1)}>  
= \nonumber \\
 -\frac{4}{9 \pi^2} \frac{e^2\hbar }{ m c^4} \Omega_0^3 \beta \times
    \sum_{n=1}^{\infty}
     \frac{(-1)^{n+1} \tilde{J_{n}}}
     { 1 + \Gamma^2 \Omega_0^2 n^2}, \;\; \beta \ll 1, \;\; \gamma \approx 1, \;\;\; \omega_0 \ll \Omega_0,
  \\ 
\tilde{J_{n}} \equiv \int_{0}^{\infty} dx \frac{ x^4 \exp (-\pi x)
(12 + \frac{n^2-1-x^2}{n^2}) }{[(n-1)^2 + x^2][(n+1)^2 + x^2]}. \nonumber \\
 \end{eqnarray}
 It is easy to find the values of integrals $\tilde{J}_n$. For example,
 $\tilde{J}_1 \approx 0.143823 $, $ \tilde{J}_2 \approx  0.0317016 $, $\tilde{J}_3 \approx 0.00867892 $, 
 $ \tilde{J}_4 \approx 0.0031658 $, 
 $\tilde{J}_5\approx 0.00139711$, $\tilde{J}_6\approx 0.000703877$ and so on. So alternating series is decreasing by absolute value significantly.  Then we can ignore all terms starting from the second one to have
 \begin{eqnarray}
 \label{eq:force}
  <F^{(1)}>_r + <F_1^{(1)}>_r + <F_{2}^{(1)}>_r \approx  -\frac{4}{9 \pi^2} \frac{e^2\hbar }{ m c^4} \Omega_0^3 \beta \times
       \frac{0.143823. }{ 1 + \Gamma^2 \Omega_0^2 }, \;\; \beta \ll 1, \;\; \gamma \approx 1, \;\;
       \omega_0 \ll  \Omega_0    
 \end{eqnarray}
  From the physical point of  view it means that all high harmonics with small wave lengths can be defied. The main contribution to the force comes from the first harmonic with wave length $ 2 \pi c / \Omega_0$ which is much greater than the circle radius 
 $\beta c / \Omega_0$ and therefore the oscillator size.  It justifies electric-dipole approximation used in the force calculation.\\
 \indent The total force has a negative value. So the force is directed to the center of rotation.  This feature of the force is discussed
 in the next section. 
 \section {About the Possibility of Non Relativistic Rotation of an Oscillator in Zero-Point Field, Without Any Other External Forces. } 
The force (\ref{eq:force}) has been calculated in an instantaneous inertial frame $I_{\tau} $ associated with a tetrad $ \mu_{\tau}$.
Using a Lorentz transform it is easy to determine a new tetrad $\lambda_{\tau}$, and the force in it, which has the same orientation in
3-space  as $\mu_{\tau}$ but is at rest in the lab system. For small velocities, $\beta \ll 1$, (\ref{eq:force}) in $\lambda_{\tau}$ will not change. The same is true for the acceleration of the oscillator center, and its value is
\begin{eqnarray}
a=\Omega_0 v= c \Omega_0 \beta.
\end{eqnarray}
If the force ( \ref{eq:force}) is the only averaged force acting on the oscillator center then
\begin{eqnarray}
\label{eq:newton_law}
 mc\Omega_0 \beta= <F^{(1)}>_r + <F_1^{(1)}>_r +<F_2^{(1)}>_r .
\end{eqnarray}
All quantities in this equation have the same direction, to the center of rotation, and we omit their negative sign.
 From (\ref{eq:newton_law}) follows 
\begin{eqnarray}
\Omega_0=\frac{2 m c^2}{\hbar} \times \frac{3}{4} (\;\; \frac{ c \hbar /e^2}{ 0.143823 /\pi^2 - e^2 /  c \hbar} \;\;  ) ^{0.5} 
\end{eqnarray}
for the frequency rotation without external forces  and
\begin{eqnarray}
R = \frac{c \beta}{\Omega_0}, \;\;\; \beta \ll 1
\end{eqnarray}
for the radius of rotation. \\
\indent Let us compare R with characteristic lengths of the electron, for example, electron classical radius $ r_{cl}$ and Zitterbewegung radius  $r_{zbw}$:
\begin{eqnarray}
r_{cl}= \frac{e^2}{m c^2}, \;\;\; r_{zbw}=\frac{\hbar}{2 m c}= r_{cl} \frac{1}{2} \,\frac{c \hbar}{ e^2} = \frac{r_{cl}}{2 \alpha},
\end{eqnarray}
where $\alpha$ is a fine structure constant, and $\frac{1}{\alpha}=137.04$. \\
\indent It is easy to see that in non relativistic case  with $\beta \ll 1$ radius of rotation in zero-point field
without external force should be much less than the classical electron radius: 
\begin{eqnarray}
R=r_{cl} (0.665243) \beta \ll r_{cl}.
\end{eqnarray}   
The trajectories with such small radii can not be observed experimentally because  our  semi classical approach does not work for $R \ll r_{cl}$.  The positive result, $R \gg r_{cl}$, would have contradicted our intuition and experimental observations. The issue  about possibility of stable orbits in the vacuum should and will be addressed to colored particles ( quarks) moving in a colored non abelian vacuum field. It is the subject of the next paper.\\
\indent There are also some works where a classical description of quark motion is being developed and even applied to experimental results \cite{Wong1970}, \cite{Arodz1982}, \cite{Arodz1983}, \cite{Cassing2013}. \\
 {\bf APPENDIX}
\appendix
\section{Appendix: Calculation of the Correlation Function $  < E^{(3)}(\tau_1) H^{(2)}(\tau_2)>_d. $ }
\label{section: Calculaion of the CF}
\indent Calculation of $ <E^{(z)}(\tau_1) H^{(y)}(\tau_2)>_d$
  in (\ref{eq:X_Component_Of_Force})  consists of several steps: \\
 1. Inserting  $E^{(z)}(\tau_1)$ and  $H^{(y)}(\tau_2)$ from
 (\ref{eq:field_Instantaneous}) and (\ref{eq:mff1}) to bilinear expression $ <E^{(z)}(\tau_1) H^{(y)}(\tau_2)>_d$, we obtain
  pretty complicated expression with two double sums, $\sum_{n_1}
 \sum_{n_2}$ and $\sum_{\lambda_1} \sum_{\lambda_2}$, and  
 double integral over the angles $\int do_1 \int do_2 \equiv \int d\theta_1 d\phi_1 \sin\theta_1 \int d\theta_2 d\phi_2 \sin\theta_2 $. \\  
2. Using (\ref{eq: CosSinAverage1})  and summing  over $\lambda_2$,
 $n_2$, and integrating over $\int do_2$, we come to the expression with single sums $\sum_{\lambda_1}$ and $\sum_{n_1}$, and
an integral over angles $\int do_1$.\\
3. Using summation over $\lambda_1$ and using 
(\ref{eq:lambda1}) we come to  the expression with one integral over angular
variables and one sum over $n_1$, $\sum_{n_1}\int d\theta_1 d\phi_1 \sin \theta_1$ or, after notation change, $\sum_{n}\int d\theta d\phi \sin \theta$ . 
This expression is still pretty long and depends on two parameters 
$\alpha_1$ and $\alpha_2$ defined in (\ref{eq: alpha}). \\
4. Introducing new parameters according to
\begin{eqnarray}
\label{eq: Alpha_Delta}
\alpha \equiv 1/2(\alpha_1+\alpha_2), \;\;\; \delta\equiv \alpha_2-\alpha_1=
\Omega_0 \gamma(\tau_2 - \tau_1),
\end{eqnarray}
we can change integration variables
\begin{eqnarray}
\hat{k}_x=\hat{k}_x^{\prime}\cos \alpha -
\hat{k}_y^{\prime}\sin\alpha, \;\; \hat{k}_y=
\hat{k}_x^{\prime}\sin \alpha + \hat{k}_y^{\prime}\cos \alpha,
\end{eqnarray}
which is equivalent to $\phi^{\prime}=\phi - \alpha$ 
 and keeps $do$ invariant. \\
\indent Finally we come to the expression for the CF (\ref{eq:I^{zy}_dd})
\begin{eqnarray}
<E^{(z)}(\tau_1)H^{(y)}(\tau_2)>_d=\frac{\hbar c k^4_0 }{2
\pi^2}\int do [-\hat{k}_x\gamma \cos
\frac{\delta}{2}-\hat{k}_y\gamma
\sin\frac{\delta}{2}+\hat{k}_x\hat{k}_y \beta \gamma
+\frac{1}{2}\beta \gamma \sin\delta (1+ \hat{k}_z^2)] \times
 \nonumber \\
 \times \sum_{n=0}^{\infty} n^3 \cos n F, \;\;\;
 F \equiv \delta [1-\beta \frac{\sin \delta /2}{\delta/2}\sin \theta \sin
 \phi], \;\;\;\; k_0=\Omega_0/c.
\end{eqnarray}
We omitted here upper index ``$\prime$"  and use dummy variables $\theta$, $\phi$, and $\hat{k}_i$ instead of $\theta^{\prime}$, $\phi^{\prime}$, and $\hat{k}^{\prime}_i$.\\
\indent So the CF depends on the difference $\tau_2-\tau_1$, an expected
feature, and index d means discrete spectrum of zero-point radiation due to 
the periodicity condition we use. \\
\indent
The function $f_d$ can also be written in the form:
 \begin{eqnarray}
 f_d(\tau_1-\tau_2) =2 \sum_{n=1}^{\infty}\frac{1}{\omega_n}\frac{\cos(\omega_n(\tau_1-\tau_2)+\phi_n)}{\sqrt{(1-\frac{\omega_0^2}
 {\omega_n^2})^2 +\Gamma^2 \omega_n^2(1+\frac{a^2}{\omega_n^2c^2})^2}}
 \end{eqnarray}
  The cosine and sine functions are defined as follows
\begin{eqnarray}
\label{eq:sin_phi_n}
\cos \varphi_n =\frac{\omega_n \Gamma (1 + \frac{a^2}{\omega_n^2 c^2})}{\sqrt{(1-\frac{\omega_0^2}{\omega_n^2})^2 +\Gamma^2 \omega_n^2(1 + \frac{a^2}{\omega_n^2 c^2})^2}},\;\;\;
\sin \varphi_n =\frac{1-\frac{\omega_0^2}{\omega_n^2}}{\sqrt{(1-\frac{\omega_0^2}{\omega_n^2})^2 +\Gamma^2 \omega_n^2(1 + \frac{a^2}{\omega_n^2 c^2})^2}}.
\end{eqnarray}
The function $f_d$ is simplified significantly if $\omega_0 \ll \Omega_0$. Then 
\begin{eqnarray}
\label{f_d}
f_d(\tau_1 - \tau_2) = \frac{2}{\Omega_0 \gamma} 
\sum_{n=1}^{\infty} \frac{1}{n} \; 
\frac{ n  ( \Gamma \Omega_0 \gamma)   \cos ( n \delta) + \sin (n \delta)
}{ 1 + (\Gamma \Omega_0 \gamma)^2 n^2 (1 + \frac{\beta^2 \gamma^2}{n^2})^2}, 
\;\;\; \delta=\Omega_0 \gamma (\tau_2-\tau_1), \;\;\; \omega_0 \ll 
\Omega_0.                                                                                                       
\end{eqnarray}
\section{Appendix: Simplification of the Renormalized Correlation Function $<E^{(z)}(\tau_1)H^{(y)}(\tau_2)>_r $}
\label{sec:SimplificationOfRenCorFunct}
Using dimensionless variables in integral over $\omega$ in expression (\ref{eq:I^{zy}_d}), more convenient
for calculations, we have
 \begin{eqnarray}
 <E^{(z)}(\tau_1)H^{(y)}(\tau_2)>_r=\frac{\hbar c k^4_0}{2
 \pi^2}
 \int do [-\hat{k}_x\gamma \cos
 \frac{\delta}{2}-\hat{k}_y\gamma
 \sin\frac{\delta}{2}+\hat{k}_x\hat{k}_y \beta \gamma
 +\frac{1}{2}\beta \gamma \sin\delta (1+ \hat{k}_z^2)] \times  \nonumber \\
 \int_0^{\infty} dx \frac{2 x^3 \cosh(x
 \delta [1-\beta \frac{\sin \delta /2}{\delta/2}\sin \theta \sin
  \phi])}{e^ {2\pi x}-1}, \\
  \delta = \Omega_0 \gamma (\tau_2- \tau_1).  \nonumber   
 \end{eqnarray} 
Changing the order of integration,  it easy to see that 
 \begin{eqnarray}
 \int_{0}^{2 \pi} d \phi \hat{k}_x \cosh( x \delta [1-\beta \frac{\sin \delta /2}{\delta/2}
 \sin \theta \sin \phi]) =0, \nonumber \\
 \int_{0}^{2 \pi} d \phi \hat{k}_y \cosh( x \delta [1-\beta \frac{\sin \delta /2}{\delta/2}
  \sin \theta \sin \phi]) =0, \nonumber \\
  \int_{0}^{2 \pi} d \phi \hat{k}_x \hat{k}_y\cosh( x \delta [1-\beta \frac{\sin \delta /2}{\delta/2}
   \sin \theta \sin \phi]) =0, \nonumber \\
   \int_{0}^{2 \pi} d \phi \cosh( x \delta [1-\beta \frac{\sin \delta /2}{\delta/2}
    \sin \theta \sin \phi]) =\cosh (x \delta)\int_{0}^{2 \pi} d \phi \cosh(2 x  \beta \sin \delta /2
        \sin \theta \sin \phi), \nonumber \\
 \end{eqnarray}
 and  
 \begin{eqnarray}
  <E^{(z)}(\tau_1)H^{(y)}(\tau_2)>_r=\frac{\hbar c k^4_0}{2
  \pi^2} \times \frac{1}{2} \beta \gamma \sin \delta   \int_0^{\infty} dx \frac{2 x^3 \cosh(x
  \delta) )}{e^ {2\pi x}-1} \int_{0}^{\pi}d \theta \sin \theta (1 + \cos^2 \theta ) \times \nonumber \\
   \int_{0}^{2 \pi}
  d \phi \cosh(2 x \beta \sin (\delta/2 ) \sin \theta \sin \phi)  \nonumber
  \end{eqnarray}
  This expression is used in (\ref{eq:RenormCorFunc})
  
  \end{document}